\documentclass[aps,prd,twocolumn,showpacs,superscriptaddress,groupedaddress,showpacs]{revtex4}  
\usepackage{graphicx}  
\usepackage{dcolumn}   
\usepackage{bm}        
\usepackage{amssymb}   

\usepackage{tikz}
\usetikzlibrary{matrix}

\usepackage{latexsym}
\usepackage{epsfig}
\usepackage{graphicx}
\usepackage{epstopdf}
\usepackage{amssymb}
\usepackage{amsmath}
\usepackage{dcolumn}
\usepackage{bm}
\usepackage{color}
\usepackage{comment}

\usepackage{float}

\graphicspath{{pictures/}}

\begin{document}

\title{\bf Accretion Processes for General
    Spherically Symmetric Compact Objects}

\author{Sebastian Bahamonde}
\email{sebastian.beltran.14@ucl.ac.uk}\affiliation{Department of Mathematics,University College London,
Gower Street, London, WC1E 6BT, UK}

\author{Mubasher Jamil}
\email{mjamil@sns.nust.edu.pk}\affiliation{Department of Mathematics, School of Natural
Sciences (SNS), National University of Sciences and Technology
(NUST), H-12, Islamabad, Pakistan}

\begin{abstract}
{\centering  \textbf{Abstract:} We investigate the accretion process for different spherically symmetric space-time geometries for a static fluid. We analyse this procedure using the most general black hole metric ansatz. After that, we examine the accretion process for specific spherically symmetric metrics obtaining the velocity of the sound during the process and the critical speed of the flow of the fluid around the black hole. In addition, we study the behaviour of the rate of change of the mass for each chosen metric for a  barotropic fluid. } 
\end{abstract}

 \maketitle

\newpage

\section{Introduction}

About two decades ago, astronomical observations such as CMB radiation \cite{Spergel:2006hy}, Supernova Type Ia \cite{Perlmutter:1998np} and large scale structure data \cite{Eisenstein:2005su,Riess:1998cb} showed that the Universe is undergoing an accelerating expansion period. This discovery was a revolution in cosmology and the agent responsible for this effect was named dark energy. This energy has the strange property that produces repulsive gravitational effects and it violates the null and weak energy condition \cite{Johri:2003rh,Lobo:2005us}. Surprisingly, observations suggests that around two-thirds of the total energy of the Universe belongs to the dark energy \cite{Spergel:2006hy,Perlmutter:1998np}. However, the nature of this energy is not well-understood and nowadays is one of the most challenging problems in theoretical physics. Over the last two decades, a number of theorists have been trying to tackle this important problem. Several ideas have been proposed like cosmological constant, phantom energy, quintessence, k-essence, dynamic scalar fields, and others. Usually, dark energy is modelled using a perfect fluid such as the pressure and energy density are related by a perfect fluid with a barotropic equation $p=w\rho$ where $w$ is the state parameter with $w=-1$ a cosmological constant, $-1<w<-1/3$ for quintessence and $\omega <-1$ for phantom models \cite{review}.

Accretion is the process by which a massive astrophysical object such as a black hole or a star can take particles from a fluid from its vicinity which leads to increase in mass (and possibly angular momentum) of the accreting body \cite{accretion}. It is one of the most ubiquitous processes in the Universe. Indeed the stars and planets form only as a result of accretion in some inhomogeneous regions of gas and dust. Existence of supermassive black holes at the centres of giant elliptical and spiral galaxies suggests that such black holes could have evolved via accretion process. Other processes of formation of giant black holes such as merger of several small mass black holes (or compact objects) or stellar collapse of several stars in a small domain leading to merger seem very remote possibilities.  The most likely model of formation of massive giant black holes is the accretion of dust or matter from nearby regions for sufficiently longer times. The formation of giant elongated astrophysical jets from active galaxies or small compact objects indicates the existence of large amounts of hot dust (likely to be in the form of accretion disk) around the regions of jet formation. However accretion process not always increases the mass of the compact source, sometimes the infalling matter is thrown away in the form of jets or cosmic rays. It is more probable that the accretion process may not be static and the velocity of free-fall and the energy density of the fluid change with time and position. The accretion of normal matter and dust onto compact objects is a well-studied problem, however the accretion of more exotic types of energy-matter is not so commonly probed including dark energy and stiff fluids. Since the Universe is dominated by dark energy, it is more pertinent to study the accretion of various forms of dark energy onto black holes.

Bondi investigated accretion for compact objects using Newtonian gravity \cite{Bondi:1952ni}. After the arrival of general relativity, it became possible to investigate the accretion onto more compact objects such as neutron stars and black holes. Michel was the first who studied the relativistic accretion process for the Schwarzschild black hole \cite{Michel:1972}. Babichev et. al derived the fate of a stationary uncharged black hole in the phantom energy dominated universe \cite{Babichev:2004yx}. They concluded that the phantom energy falling onto a black hole will decrease its mass. We use the same formalism in order to study accretion process for different spherically symmetric spacetimes. More recently, Debnath \cite{Debnath:2015hea} presented a framework of static accretion onto general static spherically symmetric black holes, thereby generalising the work of Babichev et al \cite{Babichev:2004yx}. We extend these previous studies using a more general ansatz for a static spherically symmetric spacetime which covers all possible static black hole (and non-black hole) solutions such as Schwarzschild-monopole, string cloud, Jannis-Newman-Winicour and dilaton (stringy charged) black hole. In particular, we investigate the parameters of fluid and its effect on the black hole mass.

This paper uses the geometrized units $c=G=\hbar=1$ and the signature $(-,+,+,+)$ for the metrics. This paper is organised as follows: In the second section, we derive a general formalism for the relativistic spherical accretion process. In the third section, we study some specific metrics such as black holes with topological defects, string cloud, JNW metric and a static charged black hole. For each case, we derive the velocity profiles, the speed of sound, the critical speed of the flow and the rate of change of the black hole mass for a barotropic fluid. We explain these parameters with the help of figures. Finally, we discuss the results of our paper.

\section{General formalism}

We consider the following metric ansatz for the general static spherically symmetric spacetime
\begin{eqnarray}
ds^{2}&=-A(r)dt^2+\displaystyle\frac{1}{B(r)}dr^2+C(r)(d\theta^2+\sin^2\theta d\phi^2),\,\,
\end{eqnarray}
where $A(r)>0$, $B(r)>0$ and $C(r)>0$ are functions of $r$ only. This metric ansatz with $C(r)=r^2$ was used by Debnath \cite{Debnath:2015hea}. We are going to perform a general accretion formalism for those space-times geometries.

Assuming the energy matter as a perfect fluid which is isotropic though inhomogeneous, the later assumption leads to variations in energy density at different positions near the accreting source.  The energy-momentum tensor is given by
\begin{eqnarray}
T_{\mu\nu}&=(\rho+p)u_{\mu}u_{\nu}+p\, g_{\mu\nu},
\end{eqnarray}
where $\rho$ is the energy density, $p$ the pressure and $u^{\mu}$ is the four-velocity that in general will be
\begin{equation}
u^{\mu}=\frac{dx^{\mu}}{d\tau}=(u^{t},u^{r},0,0),
\end{equation}
where $\tau$ is the proper time. We have that $u^\theta=0$ and $u^\phi=0$ due to the restriction of spherical symmetry. Note that all components of 4-velocity, pressure and energy density are functions of $r$ only. Since the 4-velocity must satisfy the normalization condition $u_{\mu}u^{\mu}=-1$ we find
\begin{eqnarray}
u^{t}:=\frac{dt}{d\tau}&=&\sqrt{\frac{u^2+B}{AB}},
\end{eqnarray}
where for simplicity we have named $u=u^{r}=\frac{dr}{d\tau}$. Due to the presence of square root, there are two possibilities: $u^t>0\ (<0)$, which respectively imply forward (backward) in time conditions. The former condition is necessary to preserve causality in the process: fluid must enter the black hole but not leave it. Moreover, to study accretion, we require $u<0$ while for any outward flows $u>0$. In the astrophysical context both inward and outward flows are important. The former ones lead to growth of black holes while the later ones lead to jets. Using the energy-momentum conservation law: $0=T^{\mu\nu}_{~~;\mu}=\frac{1}{\sqrt{-g}}(\sqrt{-g}T^{\mu\nu})_{,\mu}+\Gamma^{\nu}_{~\alpha\mu}T^{\alpha\mu},$ we find
\begin{eqnarray}
(\rho+p)u\frac{A(r)}{B(r)}\sqrt{u^2+B(r)}C(r)&=&A_{1},\label{1}
\end{eqnarray}
where $A_{1}$ is a an integration constant.

By projecting the conservation law onto the 4-velocity $u_\mu T^{\mu\nu}_{~~;\nu}=0,$ thereby contracting all indices, we can find the relativistic energy flux (or continuity) equation
\begin{eqnarray}
u^{\mu}\rho_{,\mu}+(\rho+p)u^{\mu}\,_{;\mu}&=&0.
  \end{eqnarray}
In our case, we will assume that the pressure and energy density are related by a certain equation of state $p=p(\rho)$. The last equation after simplification yields
\begin{eqnarray}
\frac{\rho'}{\rho+p}+ \frac{u'}{u}+\frac{A'}{2A}+\frac{B'}{2B}+\frac{C'}{C}&=&0.\label{rhor}
\end{eqnarray}
Here, prime denotes differentiations with respect to $r$.

Integration of (\ref{rhor})  yields
\begin{eqnarray}
uC(r)\sqrt{\frac{A(r)}{B(r)}}e^{\int{\frac{d\rho}{\rho+p(\rho)}}}&=&-A_{0},
\end{eqnarray}
where $A_{0}$ is an integration constant, while a negative sign is introduced on the right hand side since $u<0$ on the left hand side. Now, if we combine the above equation with (\ref{1}) we obtain

\begin{eqnarray}
 (\rho+p)\sqrt{u^2+B(r)}\sqrt{\frac{A(r)}{B(r)}}e^{-\int{\frac{d\rho}{\rho+p(\rho)}}}=-\frac{A_{1}}{A_{0}}\equiv A_{3},\label{ksks}
\end{eqnarray}
where $A_{3}$ is a constant which depends on $A_{1}$ and $A_{0}$. Due to spherical symmetry, we take $\theta=\pi/2$ (fluid flow in the equatorial plane) which yields, $\sqrt{-g}=C\sqrt{A/B}$. Furthermore, the equation of mass flux, $0=J^{\mu}_{~;\mu}=\frac{1}{\sqrt{-g}}\frac{d}{dr}(J^{r}\sqrt{-g})$ leads to
\begin{align}
    \rho u\sqrt{\frac{A(r)}{B(r)}}C(r)&=A_{2},\label{6}
\end{align}
where $A_{2}$ is an integration constant. If we divide Eq. (\ref{1}) with (\ref{6}), we get another useful relation
\begin{align}
\frac{(\rho+p)}{\rho}\sqrt{\frac{A(r)}{B(r)}}\sqrt{u^2+B(r)}&=\frac{A_{1}}{A_{2}}\equiv A_{4}, \label{ddjdjdj}
\end{align}
where $A_4$ is another arbitrary constant. 

Taking differentials of Eqs. (\ref{6}) and (\ref{ddjdjdj}) and solving together, we obtain
\begin{eqnarray}
\Big(V^2-\frac{u^2}{u^2+B}\Big)\frac{du}{u}+\Big((V^2-1)\Big(\frac{A'}{A}-\frac{B'}{B}\Big)&&\nonumber\\
+\frac{C'}{C}V^2-\frac{B'}{2(u^2+B)}\Big)dr=0,&&\label{lkj}
\end{eqnarray}
where we have introduced the variable
\begin{align}
V^2\equiv\frac{d \ln(\rho+p)}{d \ln\rho}-1.
\end{align}
It is important to mention that in Eq. (9) of Ref. \cite{Debnath:2015hea} a similar expression with $C(r)=r^2$ was derived which contained some erroneous signs.\\
By taking the two brackets in (\ref{lkj}) equal to zero, we can find the critical point of accretion located at $r=r_{c}$. Thus, at the critical point we have
\begin{eqnarray}
V_{c}^2&=&\frac{u_{c}^{2}}{u^{2}_{c}+B(r_{c})},\\
(V_{c}^2-1)\Big[\frac{A'(r_{c})}{A(r_{c})}-\frac{B'(r_{c})}{B(r_{c})}\Big]
&&\nonumber\\
+\frac{C'(r_{c})}{C(r_{c})}V_{c}^2&=&\frac{B'(r_{c})}{2(u_{c}^2+B(r_{c}))}.\nonumber\\
\end{eqnarray}
Here every function is evaluated at $r=r_{c}$ and $u_{c}$ is the critical speed of the flow (velocity of the flow at the critical point). We can decouple the quantities $u_c^2$ and $V_c^2$ and obtain
\begin{eqnarray}
u_{c}^{2}&=&\frac{B(r) C(r) A'(r)}{2 A(r) C'(r)},\label{uc}\\
V_{c}^2&=&\frac{C(r) A'(r)}{C(r) A'(r)+2 A(r) C'(r)}.\label{Vc}
\end{eqnarray}
The speed of sound is found to be
\begin{eqnarray}
c_{s}^{2}&=&\frac{dp}{d\rho}\Big|_{r=r_{c}}=A_{4}\sqrt{\frac{B(r_{c})}{A(r_{c})(u^{2}_{c}+B(r_{c}))}}-1.\label{sound}
\end{eqnarray}
Note that $u_c^2$ and $V_{c}^{2}$ cannot be negative, hence
\begin{eqnarray}
\frac{A'(r_{c})}{C'(r_{c})}>0.\label{condition}
\end{eqnarray}
From Eq. (\ref{condition}), it follows that for given metric components, one can find the critical radius in the given spacetime.

The rate of change of the mass of the black hole $\dot M=4\pi r^2T^r_0,$ in the accretion process for a fluid around the black hole is given by \cite{Debnath:2015hea}
\begin{eqnarray}
\dot{M}_\text{acc}&=&4\pi A_3 M^2(\rho+p),\label{Mdot}
\end{eqnarray}
where dot represents derivative with respect to time. This equation can be found by integrating the flux of the fluid over the surface of the black hole. We can see that the mass of the black hole will increase for any fluid such as $\rho+p>0$ which accretes outside the black hole. On the other hand, if the fluid is a phantom dark energy $\rho+p<0$, then the mass of the black hole will decrease. It should be emphasized that this mass of the BH that appears in the corresponding metrics (see subsequent sections), where it is generally assumed that the mass is a pure constant. However, in realistic situations, mass of BH cannot remain fixed: accretion leads to increase in mass while Hawking radiation leads to the decrease in mass. To connect the BH solutions with their astrophysics, one has to take into account the time dependence of BH mass. It is further assumed that this time dependence does not change the global geometry and the symmetries of the spacetime, therefore spacetime metrics remain static and spherically symmetric. If one is interested to consider the combined effects of accretion and evaporation from BH simultaneously, then an additional term corresponding to Hawking radiation is added to the right hand side of Eq. (\ref{Mdot}), however it is not our concern at the moment.

\section{Spherically symmetric metrics with horizons}
In this section and subsequent, we are going to discuss the case in which $A(r)=B(r)$. All the spherically symmetric metrics with horizons are included in this case. From Eqs. (\ref{uc}) and (\ref{Vc}) we have
\begin{eqnarray}
u_{c}^{2}&=&\frac{ C(r) A'(r)}{2  C'(r)},\label{ucagain}\\
V_{c}^2&=&\frac{C(r) A'(r)}{C(r) A'(r)+2 A(r) C'(r)}.\label{Vcagain}
\end{eqnarray}
Although we are focused on the BH metrics with event horizons, the present analysis is not forbidden for horizonless spacetimes. In many cases, we are concerned with the properties of fluid flow (including critical velocity, sound speed in the fluid; evolution of fluid's energy density; speed of radial flow, etc.) and the change in mass of the accreting object and therefore the horizon does not directly involve anywhere. With this perspective, one can investigate the accretion phenomenon for more exotic objects such as naked singularity and wormholes. Since the mass of BH is connected with the size of the horizon, the horizon expands due to accretion as well.

\subsection{Schwarzschild de Sitter Black Hole with a Topological Defect (Monopole) }

Dadhich et al. derived the metric of a Schwarzschild black hole with global monopole charge by relaxing asymptotic flatness and investigated several astrophysical aspects \cite{dadi}. However in the presence of cosmological constant, the metric coefficients of this geometry become modify as follows \cite{Han:2007zza}
\begin{eqnarray}
    A(r)=B(r)&=&1-\frac{2M}{r}-\frac{\lambda}{3}r^2,\label{AB1}\\
        C(r)&=&r^2(1-\eta^2),\label{C1}
    \end{eqnarray}
where $\eta$ and $\lambda$ are the parameter related to the scale of symmetry breaking and the cosmological constant respectively. This metric has a topological defect due to the deficit solid angle around the scale of a gauge-symmetry breaking. Amani and Farahani \cite{fari} investigated the phantom energy accretion onto Schwarzschild AdS black hole with global monopole and calculated the critical velocities of the flow.
For this metric we find that
\begin{eqnarray}
u_{c}^{2}&=&\frac{3 M-\lambda  r_{c}^3}{6 r_{c}},\\
V_{c}^2&=&\frac{\lambda  r_{c}^3-3 M}{9 M+3 \lambda  r_{c}^3-6 r_{c}}.
\end{eqnarray}
The condition (\ref{condition}) yields
\begin{align}
\frac{\lambda  r_{c}^3-3 M}{3 \left(\eta ^2-1\right) r_{c}^3}>0.
    \end{align}
Using (\ref{sound}), the speed of sound will be
\begin{align}
c_{s}^{2}&=A_{4} \sqrt{\frac{2r_{c}}{2 r_{c}-3 M-\lambda  r_{c}^3}}-1.
\end{align}
It is possible to integrate the conservation laws and obtain analytical expressions of the physical parameters. Here, for simplicity we will again study the barotropic case where the fluid has an equation of state such as $p(r)=w \rho(r)$.  Using  (\ref{1}), (\ref{ksks}) and (\ref{ddjdjdj}) we can find directly one set of solutions
\small{\begin{align}
\rho(r)&=\frac{\sqrt{3} A_{0} A_{2}}{\left(\eta ^2-1\right) r^{3/2} \sqrt{A_{0}^2 \left(6 M+\lambda  r^3-3 r\right)+3 A_{2}^2 A_{4}^2 r}},\label{eq1}\\
p(r)&=\frac{\sqrt{3} A_{0} (A_{0}-A_{2})}{\left(\eta ^2-1\right) r^{3/2} \sqrt{A_{0}^2 \left(6 M+\lambda  r^3-3 r\right)+3 A_{2}^2 A_{4}^2 r}},\label{eq2}\\
u(r)&=-\frac{\sqrt{A_{0}^2 \left(6 M+\lambda  r^3-3 r\right)+3 A_{2}^2 A_{4}^2 r}}{\sqrt{3} A_{0} \sqrt{r}}.\label{eq3}
\end{align}}

\begin{figure}[H]
        \includegraphics[width=8cm]{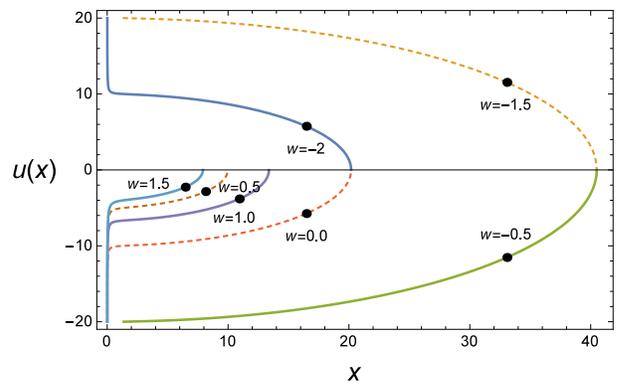}
        \caption{Velocity profile versus $x=\frac{r}{M}$ for $A_{4}=10$, $\lambda=-0.73$, $M=1$ and different values of the state parameter. For each figure, the critical radius are marked with a point for every chosen state parameter.} \label{utop}
    \end{figure}

\begin{figure}[H]
	\centering
	\includegraphics[width=8cm]{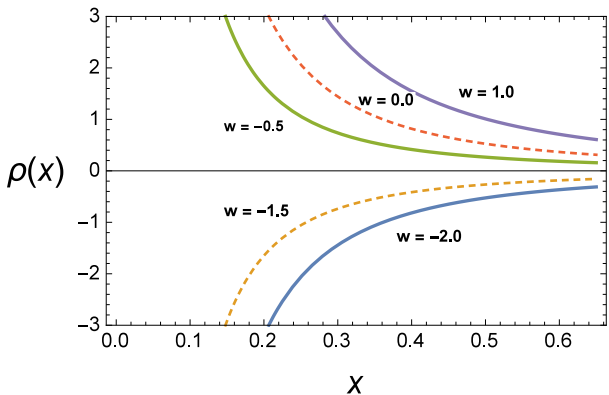}
	\caption{Energy density versus $x=\frac{r}{M}$ for $A_{2}=1$, $A_{4}=10$, $\lambda=-0.73$, $M=1$, $\eta=0.5$ and for different values of the state parameter.} \label{rhotop}
\end{figure}

\normalsize The other set of solutions are unphysical solution because they have $\rho<0$ for $w>-1$. Now, if we only use (\ref{1}) and (\ref{ddjdjdj}) and we assume a barotropic fluid such as $p=w\rho$ we obtain
\scriptsize{\begin{align}
\rho(r)&=\frac{\sqrt{3} A_{2} (w+1)}{\left(\eta ^2-1\right) r^{3/2} \sqrt{r \left(3 A_{4}^2+(w+1)^2 \left(\lambda  r^2-3\right)\right)+6 M (w+1)^2}},\label{rhorho}\\
u(r)&=-\frac{\sqrt{r \left(3 A_{4}^2+(w+1)^2 \left(\lambda  r^2-3\right)\right)+6 M (w+1)^2}}{\sqrt{3} \sqrt{r} (w+1)}.\label{uu}
\end{align}}
\normalsize  Note that if we compare both solutions, we can notice that if we replace $A_{0}=1+w$ and $A_{2}=1$ in (\ref{eq1})-(\ref{eq3}) we obtain the same solution as the equations above. Therefore, if we choose $A_{0}=1+w$ and $A_{2}=1$ there is an equivalence between solving the system of equations given by (\ref{1}), (\ref{ksks}) and (\ref{ddjdjdj}) and solving the system of equation given by (\ref{1}), (\ref{ddjdjdj}) and $p=w\rho$. Thus, in order to study the accretion process for different kind of barotropic fluids, we will choose $A_{0}=1+w$ and $A_{2}=1$. Using this argument, for the definitions of the constants we must have that $A_{3}=A_{2}A_{4}/(1+w)$. This interesting fact is because (\ref{ksks}) is just a contraction of (\ref{1}), and therefore there are not more degrees of freedom in our equations and it seems like the only possibly solution for the system is to have a barotropic fluid (with different possibles state parameters).  For all the following subsections, it can be showed that there is also an equivalence too.\\
Fig.~(\ref{utop}) represents the absolute value of the velocity profile for different values of the state parameter. Here $w<-1$, $-1<w<-1/3$, $w=-1$, $w=0$, $w=1$ refer to phantom energy, quintessence, cosmological constant, dust and stiff matter, respectively. The dots denote the critical velocity of the fluid at a certain critical radius. The behaviour shown in these figures is consistent with previous results reported in \cite{Ganguly:2014cqa,Babichev:2014lda}. It can be observed that when $w\geq0$ increases, the location of critical radius shifts to the left. Thus the infalling fluid acquires supersonic speeds closer to the black hole. We also took $\lambda<0$ in figures, which correspond to anti-de Sitter background.
Fig~(\ref{rhotop}) represents the behaviour of the energy density of fluid for different values of the state parameter. It can be seen that the energy density of the fluid decreases as it approaches the black hole when $w=-2$ and $w=-1.5$ while it increases in the remaining cases. In the former cases, the energy density becomes negative which violate the energy conditions too. However, in absolute value, the energy density always increases for all cases. Asymptotically the fluid energy density approaches zero at infinity while it approaches to a maximum near the black hole due to the strong (and perhaps quantum) gravitational interaction.

From (\ref{Mdot}), the rate of change of the mass of the black hole due to the accretion process for a barotropic fluid becomes
\begin{eqnarray}
    \dot{M}&=-\frac{4 \sqrt{3} \pi A_{2}^{2} A_{4} M^2 (w+1)}{\left(\eta ^2-1\right) r^{3/2} \sqrt{r \left(3A_{4}^2+(w+1)^2 \left(\lambda  r^2-3\right)\right)+6 M (w+1)^2}}.\label{mdotdot}
\end{eqnarray}

\begin{figure}[H]
        \centering
        \includegraphics[width=8cm]{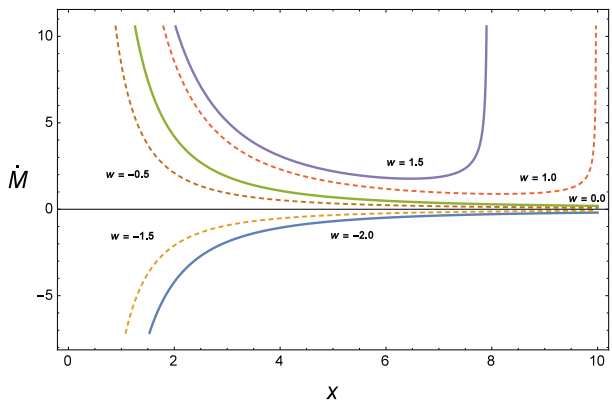}
        \caption{Rate of change of the mass of the black hole versus $x=\frac{r}{M}$ for a barotropic fluid for different values of the state parameter, $A_{2}=1$, $A_{4}=10$,  $M=1$, $\lambda=-0.73$ and $\eta=0.5$.} \label{Mtop}
    \end{figure}

Fig (\ref{Mtop}) represents the rate of the change of mass of the black hole for different values of the state parameter and same initial black hole mass. For small values of $x$, near the black hole, $\dot M>0$ showing that mass of black hole increases for $w=0$ and $w=-0.5$ being matter and quintessence respectively. In other words, mass of the black hole will increase for matter and quintessence accretions. However for stiff matter accretion, $\dot M>0$ for both small and large values of $x$, thereby increasing in mass of black hole. It should be noted that quintessence, dust and stiff matter satisfy the relativistic energy conditions  for energy-matter which ensure the increase in BH mass and satisfy the second law of black hole thermodynamics. It is also apparent that $\dot M<0$ for phantom-like equations of state such as $w=-1.5,-2$, thus mass of BH will decrease by the accretion of phantom like fluids.

It is important to remark that we have another possibility  to study the accretion process for this metric, which is by taking $A_{2}<0$ and using (\ref{rhorho}) and (\ref{uu}) with a minus of difference. In that case  we have that $\rho>0$ for $w>-1$ as we required but $u(r)<0$ for $w<-1$. However, the case $A_{2}<0$ seems to be an unnatural choice since we can notice that we will need a barotropic case such as $p=-w \rho$ to have identical solutions to solving (\ref{1}), (\ref{ksks}) and (\ref{ddjdjdj}) directly and setting $A_{0}=1+w$ and $A_{2}=-1$. In other words, it is the same to have $A_{0}=1+w$ and $A_{2}=-1$ and solving the system of equations (\ref{1}), (\ref{ksks}) and (\ref{ddjdjdj}) than solving the system of equations (\ref{1}),(\ref{ddjdjdj}) and $p=-w \rho$. Therefore, we will have the same behaviour as we discussed above since for $A_{2}=-1$, the state parameter $w$ will be minus the state parameter in the case of $A_{2}=1$. Thus, for all the following subsections, we are going to find the solutions using (\ref{1}),(\ref{ddjdjdj}) and $p=w \rho$ and we will take $A_{2}=1$.

\subsection{Schwarzschild Black Hole in a String Cloud Background}

The metric for a string cloud arises from the Nambu-Goto action of a string evolving in spacetime. In 4-dimensions, the metric of a Schwarzchild black hole in a string cloud background is given by \cite{Ganguly:2014cqa}
\begin{eqnarray}
A(r)=B(r)&=&1-\frac{2M}{r}-\alpha,\label{AB2}\\
C(r)&=&r^2,\label{C2}
\end{eqnarray}
where $\alpha$ is the string cloud parameter which is $\alpha\neq1$ and $M$ is the mass of black hole and it appears as a constant of integration during the solution of field equations \cite{let}. The limit $\alpha=0$ gives the Schwarzschild solution.
The radial velocity and the energy density for a barotropic fluid are given by
\begin{align}
u(r)=\frac{\sqrt{r \left(A_{4}^2+(\alpha -1) (w+1)^2\right)+2 M (w+1)^2}}{\sqrt{r} (w+1)}\label{u22},\\
\rho(r)=\frac{A_{2} (w+1)}{r^{3/2} \sqrt{r \left(A_{4}^2+(\alpha -1) (w+1)^2\right)+2 M (w+1)^2}}.\label{rhostringg}
\end{align}
Exactly as before, we have two set of solutions for $u(r)$ and $\rho(r)$. For this case, the critical values are given by
\begin{eqnarray}
r_{c}&=&-\frac{3 M (w+1)^2}{2 \left(A_{4}^2+(\alpha -1) (w+1)^2\right)},\\
u_{c}^{2}&=&\frac{M}{2r_{c}},\\
V_{c}^2&=&-\frac{M}{3 M+2 (\alpha -1) r_c}.
\end{eqnarray}
As a special case, if $A_4=1+w$, then the above equations yield $r_c=-3M/1+2\alpha$, which for $\alpha=0$ gives the critical radius for the Schwarzschild black hole. Since $r_c$ is a function of $w$, its value will be different for every fluid.
\begin{figure}[H]
        \centering
        \includegraphics[width=8cm]{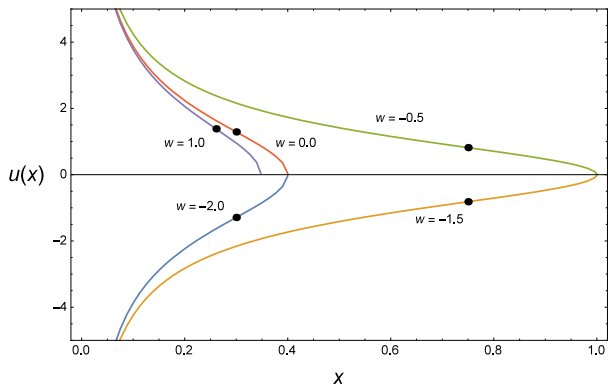}
        \caption{velocity profile versus $x=\frac{r}{M}$ for  $A_{4}=1$, $\alpha=-5$, $M=1$ and different values of the state parameter. For each figure, the critical radius are marked with a point for every chosen state parameter.} \label{ustring}
    \end{figure}
    \begin{figure}[H]
        \centering
        \includegraphics[width=8cm]{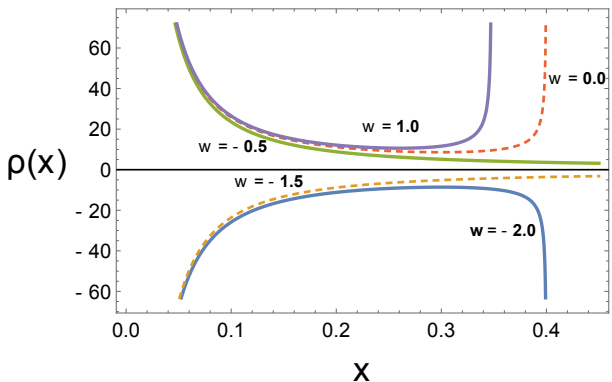}
        \caption{Energy density versus $x=\frac{r}{M}$ for $A_{2}=1$, $A_{4}=1$, $\alpha=-5$, $M=1$ and for different values of the state parameter.} \label{rhostring}
    \end{figure}
Eq. (\ref{condition}) is completely satisfied since $M>0$. The speed of sound is given by
\begin{align}
    c_{s}^{2}&=A_{4} \sqrt{\frac{1}{1-\alpha -\frac{3 M}{2 r_{c}}}}-1.
\end{align}
Thus, the rate of change of the mass of the fluid is
\small{\begin{equation}
    \dot{M}=\frac{4 \pi  A_{2}^{2}A_{4} M^2 (w+1)}{r^{3/2} \sqrt{r \left(A_{4}^2+(\alpha -1) (w+1)^2\right)+2 M (w+1)^2}}.\label{Mdotstring}
\end{equation}}
\begin{figure}[H]
    \centering
    \includegraphics[width=8cm]{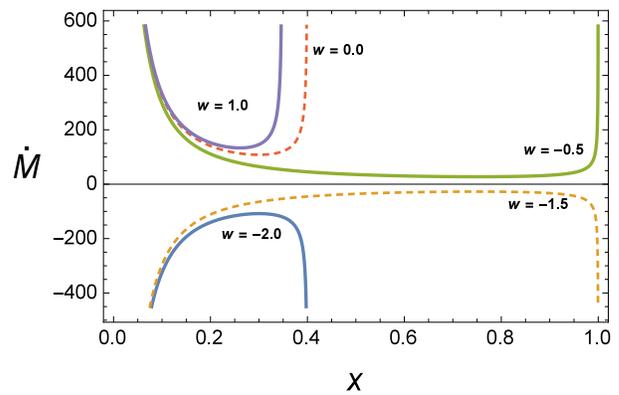}
    \caption{\normalsize Rate of change of the mass of the black hole versus $x=\frac{r}{M}$ for a barotropic fluid for different values of the state parameter, $A_{2}=1$, $A_{4}=1$, $M=1$ and $\alpha=-5$. } \label{Mstring}
\end{figure}
\normalsize From Fig. (\ref{ustring}), it can be seen that the radial velocity of the fluid is negative for the cases $w=-1.5$ and $w=-2$ while it is positive for $w=-0.5,0,1$. If the fluid flow is inwards then $u>0$ is not allowed and vice versa. In the case of quintessence, the fluid velocity is zero (at rest) at $x=1$ while it attains critical velocity  at approximately $x=0.7$. Similarly for dust and stiff matter, the critical values of velocity occur at approximately $x=0.3$. Similarly, accretion of phantom fluids lead to critical flows after $x=0.7$. In Fig.~(\ref{rhostring}), we plot the evolution of energy density of fluids in the vicinity of the black hole. It is apparent that for fluids satisfying the weak and dominant energy conditions such as dust, quintessence and stiff fluid, the energy density increases as the fluid moves towards the BH while the reverse happens for  $w=-1.5$ and $w=-2$. In Fig. (\ref{Mstring}), we make a plot of rate of change of mass of BH against $x$ and find that mass of the BH will increase due to the fluids satisfying energy conditions while it decreases for fluids violating the same energy conditions.\\

\subsection{Janis-Newman-Winicour Spacetime}

The physical existence of naked singularities is still questionable,
however there are few exact solutions of Einstein field equations
that represent naked singularity. These include the JNW,
JMN and gamma metrics \cite{joshi}. The Janis-Newman-Winicour (JNW)
solution is obtained as an extension of the Schwarzschild space-time
when a massless scalar field (with vanishing potential) is
introduced. Thus this solution is not a vacuum solution \cite{zhou}.
Here the coordinate singularity in the Schwarzschild space-time
becomes a naked singularity in JNW spacetime. Joshi and
collaborators \cite{joshi} (see also references therein) have
investigated several astrophysical features distinguishing a black
hole from a naked singularity. The gravitational lensing features
due to naked singularity have also been explored \cite{viru}.

There are various forms of JNW metric available in literature,
however we adopt the following form \cite{Chowdhury:2011aa}
\begin{eqnarray}
A(r)=B(r)&=&\Big[\frac{2r-r_{0}(\mu-1)}{2r+r_{0}(\mu+1)}\Big]^{\frac{1}{\mu}},\\
C(r)&=&\frac{1}{4}\frac{[2r+r_{0}(\mu+1)]^{\frac{1}{\mu}+1}}{[2r-r_{0}(\mu-1)]^{\frac{1}{\mu}-1}},
\end{eqnarray}
where $\mu$ is the scalar charge which is defined in the interval $(1,\infty)$ and $r_{0}$ is related to the mass by $r_0=2M$. The Schwartzschild metric can be obtained by taking $\mu=1$ after a coordinate transformation.
The scalar field is
\begin{eqnarray}
\varphi&=&\frac{a}{\mu}\ln\Big|\frac{2r-r_{0}(\mu-1)}{2r+r_{0}(\mu+1)}\Big|,
\end{eqnarray}
where $a$ and $\mu$ are related by
\begin{align}
    \mu&=1+\frac{32\pi a^{2}}{r_{0}^2}.
    \end{align}
In this case, the 4-velocity and the energy density of the barotropic fluid are

\begin{figure}[H]
        \centering
        \includegraphics[width=8cm]{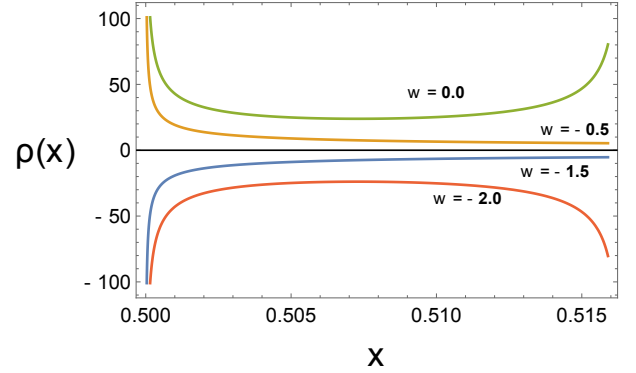}
        \caption{Energy density versus $x=\frac{r}{M}$ for $A_{2}=1$, $A_{4}=0.3$, $\mu=2$, $r_{0}=1$ and for different values of the state parameter.} \label{rhoJNWfig}
    \end{figure}

\small{\begin{align}
    u(r)&=\frac{\sqrt{A_{4}^2-(w+1)^2 (2 r-\mu  r_{0}+r_{0})^{1/\mu } (2 r+\mu  r_{0}+r_{0})^{-1/\mu }}}{w+1},\label{ujnw}\\
    \rho(r)&=\frac{4 A_{2} (w+1) (2 r-\mu  r_{0}+r_{0})^{\frac{1}{\mu }-1} (2 r+\mu  r_{0}+r_{0})^{-\frac{\mu +1}{\mu }}}{\sqrt{A_{4}^2-(w+1)^2 (2 r-\mu  r_{0}+r_{0})^{1/\mu } (2 r+\mu  r_{0}+r_{0})^{-1/\mu }}}.\label{rhoJNW}
\end{align}}
    \begin{figure}[H]
        \centering
        \includegraphics[width=8cm]{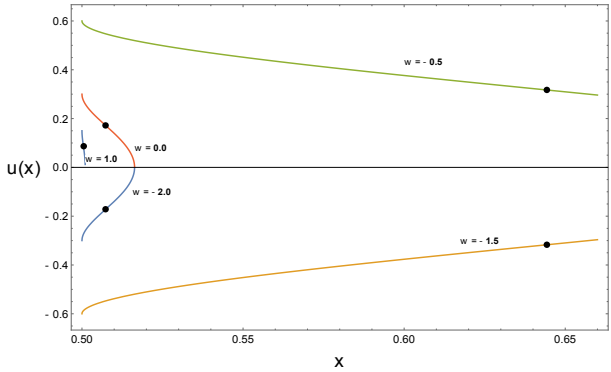}
        \caption{\normalsize Velocity profile versus $x=\frac{r}{M}$ for  $A_{4}=0.3$, $M=1$,
        $\mu=2$, $r_{0}=1$ and different values of the state parameter. For each figure,
         the critical radius are marked with a point for every chosen state parameter.} \label{uJNW}
        \end{figure}
\normalsize In Fig. (\ref{uJNW}), we plot the radial velocity of the fluid for
different values of the state parameter. The fluids have zero radial velocity in the asymptotic limit, however they acquire non-vanishing velocities as they approach the naked singularity. Due to strong gravitational attraction near the naked singularity, the fluids achieve critical velocities for different values of $x$, however interestingly we do observe some symmetry in the profile of the velocity curves and the location of critical points. After passing the critical point, the fluid flow becomes super-sonic or trans-sonic. In Fig. (\ref{rhoJNWfig}), the graph of energy density of different fluids in the neighbourhood of naked singularity. The fluid energy density for dust and quintessence rises in the near vicinity of singularity and goes to zero asymptotically. Similarly, we also see that energy density of dust and quintessence remains positive while for phantom fluids, it is negative. 

\begin{figure}[H]
    \centering
    \includegraphics[width=8cm]{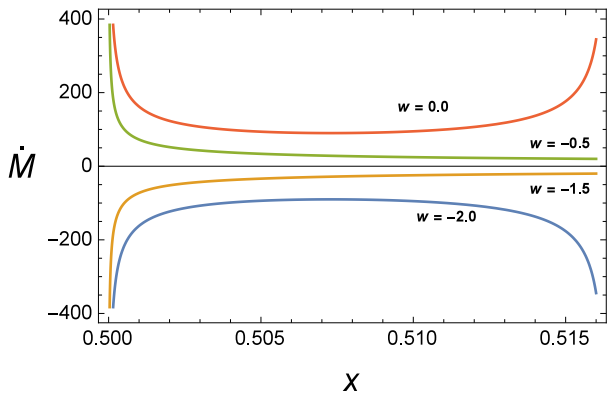}
    \caption{Rate of change of the mass of the black hole versus $x=\frac{r}{M}$ for a barotropic fluid for different values of the state parameter, $A_{2}=1$, $A_{4}=0.3$, $M=1$, $\mu=2$ and $r_{0}=1$. } \label{MJNWcomplete}
\end{figure}

In addition, for this metric we find that
\begin{eqnarray}
u_{c}^{2}&=&\frac{r_{0}}{4r_{c}} \left(\frac{2 (2 r_{c}+r_{0})}{2 r_{c}+  (\mu+1)r_{0}}-1\right)^{1/\mu },\\
V_{c}^2&=&\frac{r_{0}}{4 r_{c}+r_{0}},\\
c_{s}^{2}&=&2 A_{4} \sqrt{\frac{r_{c}}{4 r_{c}+r_{0}}}\left(\frac{2 r_{c}(1-\mu)  r_{0}}{2 r_{c}+(\mu+1)  r_{0}}\right)^{-\frac{1}{2\mu}}-1.\,\,\,\,
\end{eqnarray}
Using the condition (\ref{condition}) we obtain
\begin{align}
2 r_{0} (2 r_{c}-\mu  r_{0}+r_{0})^{\frac{1}{\mu }-1} \left(\frac{2 r_{c}-\mu  r_{0}+r_{0}}{2 r_{c}+\mu  r_{0}+r_{0}}\right)^{1/\mu }\times&\nonumber\\
(2 r_{c}+\mu  r_{0}+r_{0})^{-\frac{\mu +1}{\mu }}&>0.
\end{align}
For this case, the fluid will have a change of its mass as follows
 \scriptsize{\begin{equation}
   \dot{M}=\frac{16 \pi  A_{2}^{2} A_{4} M^2 (w+1) (2 r+(1-\mu)  r_{0})^{\frac{1}{\mu }-1} (2 r+(\mu+1)  r_{0})^{-\frac{\mu +1}{\mu }}}{\sqrt{A_{4}^2-(w+1)^2 (2 r+(1-\mu)  r_{0})^{1/\mu } (2 r+(\mu+1)  r_{0})^{-1/\mu }}}.\label{mdotdotJNW}
\end{equation}}
\normalsize From Fig. (\ref{MJNWcomplete}), that the mass of the JNW singularity increases in the cases of dust and quintessence while it decreases for phantom fluids.

\subsection{Charged Black Hole in String Theory}

\begin{figure}[H]
        \centering
        \includegraphics[width=8cm]{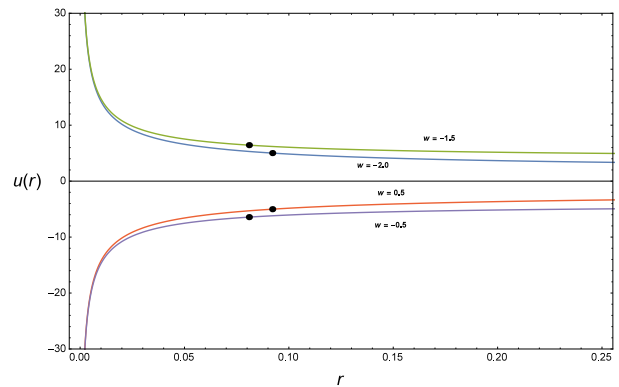}
        \caption{Velocity profile versus $x=\frac{r}{M}$ for  $A_{4}=2.1$, $M=1$, $Q=1.5$, $\Phi_{0}=-0.1$ and different values of the state parameter. For each figure, the critical radius are marked with a point for every chosen state parameter.} \label{ucharged1}
    \end{figure}

\begin{figure}[H]
        \centering
        \includegraphics[width=8cm]{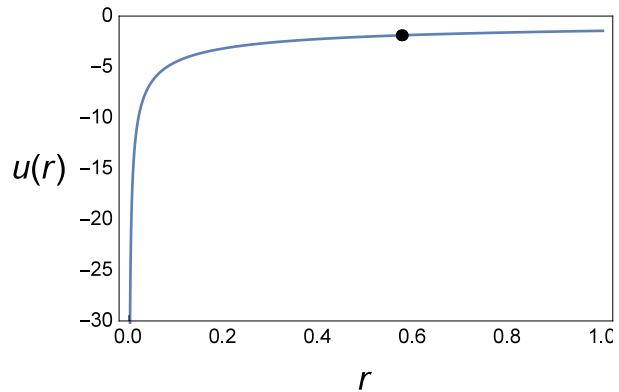}
        \caption{Velocity profile versus $x=\frac{r}{M}$ for  $A_{4}=2.1$, $M=1$, $Q=1.5$, $\Phi_{0}=-0.1$ and for $w=1.0$. The critical radius are marked with a point.} \label{ucharged2}
    \end{figure}

\begin{figure}[H]
        \centering
        \includegraphics[width=8cm]{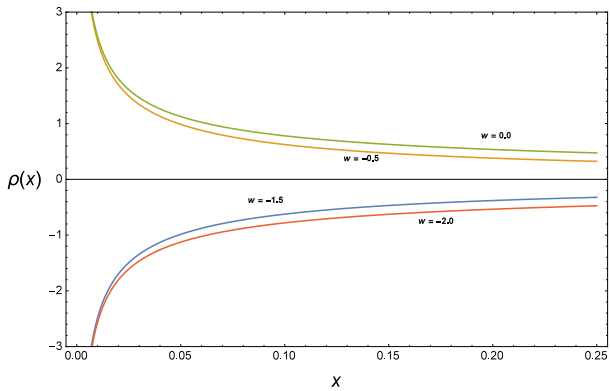}
        \caption{Energy density versus $x=\frac{r}{M}$ for  $A_{4}=2.1$, $A_{2}=1$, $M=1$, $Q=1.5$, $\Phi_{0}=-0.1$ and for different values of the state parameter.} \label{rhocharged1}
    \end{figure}

\begin{figure}[H]
        \centering
        \includegraphics[width=8cm]{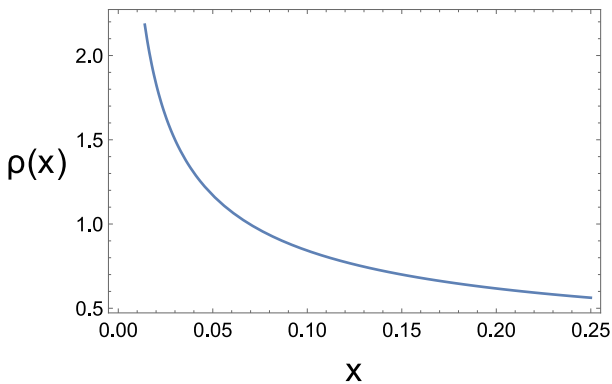}
        \caption{Energy density versus $x=\frac{r}{M}$ for  $A_{4}=2.1$, $A_{2}=1$, $M=1$, $Q=1.5$, $\Phi_{0}=-0.1$ and for $w=1$.} \label{rhocharged2}
    \end{figure}

\begin{figure}[H]
        \centering
        \includegraphics[width=8cm]{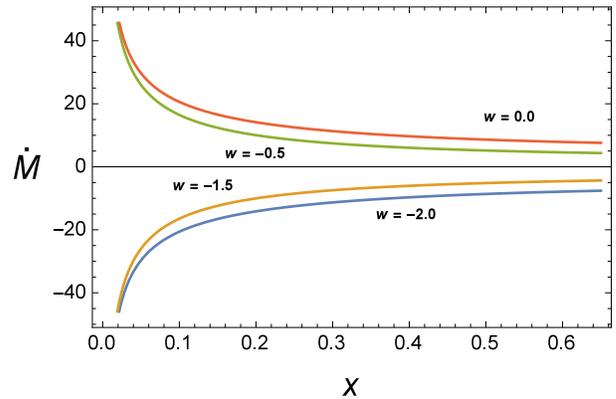}
        \caption{Rate of change of the mass of the black hole versus $x=\frac{r}{M}$ for a barotropic fluid with different values of the state parameter and the following parameters: $A_{4}=2.1$, $A_{2}=1$, $M=1$, $Q=1.5$ and $\Phi_{0}=-0.1$.} \label{Mcharged}
    \end{figure}

\begin{figure}[H]
  	\centering
  	\includegraphics[width=8cm]{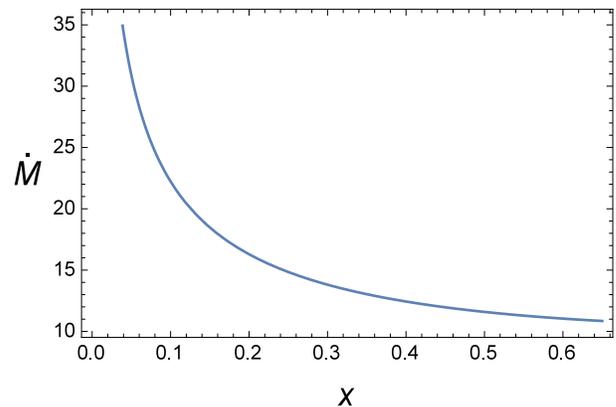}
  	\caption{Rate of change of the mass of the black hole versus $x=\frac{r}{M}$ for a barotropic fluid with $w=1$ and the following parameters: $A_{4}=2.1$, $A_{2}=1$, $M=1$, $Q=1.5$ and $\Phi_{0}=-0.1$.} \label{Mcharged2}
  \end{figure}

Gibbons and Maeda \cite{40} and later on independently
Garfinkle, Horowitz, and Strominger \cite{41} discovered a static spherical symmetric
charged black hole in low energy effective theory of heterotic
string theory in four dimensions. Sharif and Abbas \cite{sharif} investigated the dynamical behavior of phantom energy $w<-1$ near stringy charged black hole. They deduced that this mechanism reduces the mass of the black hole, though the black hole cannot be convertible to the extremal black hole. We again analyze this black hole for accretion dynamics without restricting $w$.
  We write down the stringy charged metric in the following form \cite{Gad:2003yk}
\begin{eqnarray}
A(r)=B(r)&=&1-\frac{2M}{r},\\
C(r)&=&\Big(1-\frac{Q^2\exp{(-2\Phi_{0})}}{Mr}\Big)r^2,
\end{eqnarray}
where $Q$ is the charge parameter and $\Phi_{0}$ the asymptotic value of the dilatonic field which can be set to zero under certain cases \cite{ran}.
In this case, we have
\begin{eqnarray}
u_{c}^{2}&=&\frac{M Q^2-M^2 r e^{2 \Phi_{0}}}{Q^2 r-2 M r^2 e^{2 \Phi_{0}}},\\
V_{c}^2&=&\frac{M \left(Q^2-M r e^{2 \Phi_{0}}\right)}{Q^2 (r-M)+M r e^{2 \Phi_{0}} (3 M-2 r)},\\
c_{s}^{2}&=&A_{4} \sqrt{\frac{r \left(2 M r e^{2 \Phi_{0}}-Q^2\right)}{Q^2 (M-r)+M r e^{2 \Phi_{0}} (2 r-3 M)}}-1.\,\,\,\,
\end{eqnarray}
The radial component of the 4-velocity of the fluid and the energy density are respectively
\small{
\begin{align}
u(r)=-\frac{\sqrt{r \left(A_{4}^2-(w+1)^2\right)+2 M (w+1)^2}}{\sqrt{r} (w+1)},\label{ucharged}\\
\rho(r)=-\frac{A_{2} M e^{2 \Phi_{0}} (w+1)}{\sqrt{r} \sqrt{r \left(A_{4}^2-(w+1)^2\right)+2 M (w+1)^2} \left(M r e^{2 \Phi_{0}}-Q^2\right)}.\label{rhorhocharged}
\end{align}}
\normalsize In Figs. (\ref{ucharged1}) and (\ref{ucharged2}), we plotted the radial component of the fluid flow velocity against the dimensionless radial coordinate $x$. We also represent the location of critical velocity by thick dots whose presence in the fluid flow profiles show that critical flow is a generic property of all fluids in the strong gravitational regimes. All velocity profiles $u(x)>0$ are physically forbidden since nothing can escape from the black hole. However flows of dust, stiff matter and quintessence is permissible under our assumptions of inward flows $u<0$ for all $x$.
In Figs. (\ref{rhocharged1}) and (\ref{rhocharged2}), we plotted the energy density profiles of the fluids in the vicinity black hole. For phantom-like fluids, the energy density is negative while for the    quintessence, dust and stiff fluids the energy density is positive. It also shows that fluids will have higher densities compared to their asymptotic values.
By using (\ref{condition}) we find the following constraint for the critical radius
\begin{align}
\frac{2 M^2 e^{2 \Phi_{0}}}{r^2 \left(2 M r e^{2 \Phi_{0}}-Q^2\right)}>0.
\end{align}
Therefore, the rate of change of the mass is given by
   \scriptsize{ \begin{equation}
    \dot{M}=-\frac{4 \pi A_{2}^{2}A_{4} M^3 e^{2 \Phi_{0}} (w+1)}{\sqrt{r} \sqrt{r (A_{4}-w-1) (A_{4}+w+1)+2 M (w+1)^2} \left(M r e^{2 \Phi_{0}}-Q^2\right)}.\label{mdotdotcharged}
    \end{equation}}
\normalsize In Figs. (\ref{Mcharged}) and (\ref{Mcharged2}), the rate of change in BH mass $\dot M$ is plotted against $x$ for different values of state parameter. In all cases, it is observed that the BH mass will increase by the accretion of quintessence, dust or the stiff fluid when $x$ is small. Strikingly, the maximum rate of increase in BH mass arises due to quintessence followed by dust and stiff fluid.

\section{Discussion}

We have proposed a most general framework for the study of spherical accretion onto spherically symmetric compact objects. We explored various features of the fluid flow near black holes. Although we assumed all the fluids to satisfy a certain linear equation of state with distinct state parameters, the above analysis can alternatively be performed without specifying the equation of state, since the number of equations are sufficient to close the system of equations. Interestingly we found a link between the barotropic equation of state and the conservation laws. However the equation of state helps us identify which kind of fluid is falling onto the black hole. In other words there is no multiple accretion scenario into play, though it can be done. In particular, we focused on dust, stiff matter, quintessence and phantom dark energy accretion onto black holes unlike previous studies in literature which focused on only one kind of test fluid. We did not consider the case of cosmological constant or vacuum energy since its accretion does not alter the evolution of black holes. It is found that different fluids with distinct state parameters have different evolutions in the black hole backgrounds. Certain fluids acquire positive or negative energy density near the black hole while some fluids cause black hole masses to increase or decrease. Although we plotted all the fluid behaviours via single graphs, it is assumed that a single test fluid is accreted at one time. For a future work it will be interesting to perform similar analysis with a dynamically spherically symmetric geometry  or with a non-static fluid.

\begin{center}
    \textbf{Acknowledgements}
\end{center}
S.B. is supported by the Comisi\'on Nacional de Investigaci\'on
Cient\'ifica y Tecnol\'ogica (Becas Chile Grant No. 72150066).

\end{document}